\documentclass[useAMS,usenatbib]{mn2e}
\usepackage{mathrsfs} 
\usepackage[dvipdfmx]{graphicx}
\usepackage{mediabb}
\usepackage{color}

\def\vlsr{V_{\rm LSR}} \def\deg{^\circ} 
\def\r{\bibitem[]{}}     
\def\kms{km s$^{-1}$}      
 
\def\Xhi{X_{\rm HI}}    
\def\Hcm2{ H cm$^{-2}$ } 
\def\Ihi{I_{\rm HI}}
\def\Tg{ T_{\rm G} } \def\Tcmb{T_{\rm CMB}}
\def\Ts{T_{\rm S}} 
\def\Tc{T_{\rm C}} \def\Tcon{T_{\rm C,on}} \def\Tb{T_{\rm B}}\def\Tbon{T_{\rm B,on}} \def\Taon{T_{\rm abs, on}}
\def\Hfactor{ \Gamma }

\title[Optical Thickness of HI Gas]{Optical Thickness, Spin Temperature, and Correction Factor for Density of the Galactic HI Gas}
 
\author[Y. Sofue]{Yoshiaki Sofue$^1$\thanks{Email:sofue@ioa.s.u-tokyo.ac.jp}\\
$^1${Insitute of Astronomy, The University of Tokyo, Mitaka, Tokyo 181-0015, Japan}  
}
  
\begin{document}
\date{}
\maketitle    
\begin{abstract}
 A method to determine the spin temperature of the local ($\vlsr=0$ \kms) HI gas using saturated brightness temperature of the 21-cm line in the radial-velocity degenerate regions (VDR) is presented. The spin temperatures is determined to be $\Ts= 146.2\pm 16.1$ K by measuring saturated brightness in the VDR toward the Galactic Center, $146.8\pm 10.7$ K by $\chi^2$ fitting of expected brightness distribution to observation around the VDR, and $144.4\pm 6.8$ K toward the local arm. Assuming $\Ts=146$ K, a correction factor $\Hfactor$ for the HI density, defined by the ratio of the true HI density for finite optical thickness to that calculated by assuming optically thin HI, was obtained to be $\Hfactor \sim 1.2$ (optical depth $\tau \sim 0.3$) in the local HI gas, $\sim 1.8$ ($\sim 1.3$) toward the arm and anti-center, and as high as $\sim 3.6$ ($ \sim 2.7$) in the Galactic Center direction. It is suggested that the HI density and mass in the local arm could be $\sim 2$ times, and that in the inner Galaxy $\sim 3.6$ times, greater than the currently estimated values.
\end{abstract} 

\section{Introduction}
 
The density of Galactic HI gas is usually calculated approximately with the two assumptions that the 21-cm line is optically thin, and that the background continuum emission is sufficiently weak. This approximation is convenient, because the spin temperature of HI does not appear in the conversion relation from observed HI intensity to the hydrogen volume or column density. 
However, the approximation significantly under-estimates the HI density, when the two assumptions are not valid. Precise densities can be estimated by using a general conversion relation including both the spin temperature and continuum brightness for finite optical thickness of the HI line.

The HI line emission is often observed to have brightness temperatures as high as $\Tb \sim $ several tens to $\sim 100$ K (e.g., Kalberla et al. 2003), comparable to the spin temperature, $\Ts$, indicating that the optical thickness may not be so small. Also, radio continuum emission of the galactic disk is not sufficiently weak in the inner Galaxy that it can be ignored. In such regions the continuum emission is absorbed by the HI gas, causing apparently lower HI brightness, leading to lower HI density by the thin assumption than the true value.

Thus, the optically thin assumption may not be valid in regions having high HI brightness and/or bright continuum background. Moreover, such regions are predominantly distributed in the Galactic plane, so that the HI density and mass in the Galactic disk may have been significantly under estimated.

The spin (excitation) temperature, $\Ts$, of the interstellar HI gas has been studied extensively by analyzing emission and absorption profiles of 21-cm line spectra towards individual HI clouds, molecular clouds, and/or dark clouds located in front of galactic and extragalactic radio continuum sources
(Brown et al. 2014; 
Chengalur et al. 2013; 
Dickey et al. 2003;
Fukui et al. 2014, 2015;  
Goldsmith and Li 2005;
Heiles and Troland 2003a, b; 
Kuchar and Bania 1990;  
Li and Goldsmith 2003; 
Liszt 1983, 2001; Liszt et al. 1993;
Mebold et al. 1982;
Murray et al. 2015; 
Roberts et al.1993; 
Roy et al. 2013a,b;
Stark et al. 1994; 
Wolfire et al. 1995; 
). 
The currently measured temperatures range from $\Ts \sim 20$ to $\sim 300$ K for cold HI component, and $\Ts \sim 2000$ to $\sim 10000$ K for warm component, respectively corresponding to the cold (CNM) and warm (WNM) neutral material of the two stable phases of the interstellar pressure equilibrium (Field et al. 1969). In these studies, analyses have been made by using 21-cm line profiles in the velocity (frequency) space, and the observations have been obtained along specific lines of sights in the directions of radio continuum sources.

In the present paper, we propose a method to measure the mean spin temperature of HI gas in the galactic disk, and apply it to the HI brightness distribution in the directions of velocity-degenerate regions (VDR) toward the Galactic Center (GC), anti-Center, and tangential directions to the solar circle. In these directions, especially toward the GC, the HI line intensity at $\vlsr \sim 0$ \kms is almost saturated due to the large optical depth 

Using the measured spin temperature, $\Ts$, we derive optical depth, $\tau$, of the HI line, and obtain a correction factor, $\Hfactor$, to convert the HI density calculated under optically thin assumption to a more general and reliable density for finite optical thickness. In the analyses we make use of the Leiden-Argentine-Bonn (LAB) all-sky HI survey (Kalberla et al. 2005) and the Rhodes 2300 MHz radio continuum survey (Jonas et al. 1985).

\section{Data}

\subsection{HI map}

\begin{figure} 
\begin{center}
\hskip-5mm\includegraphics[width=9cm]{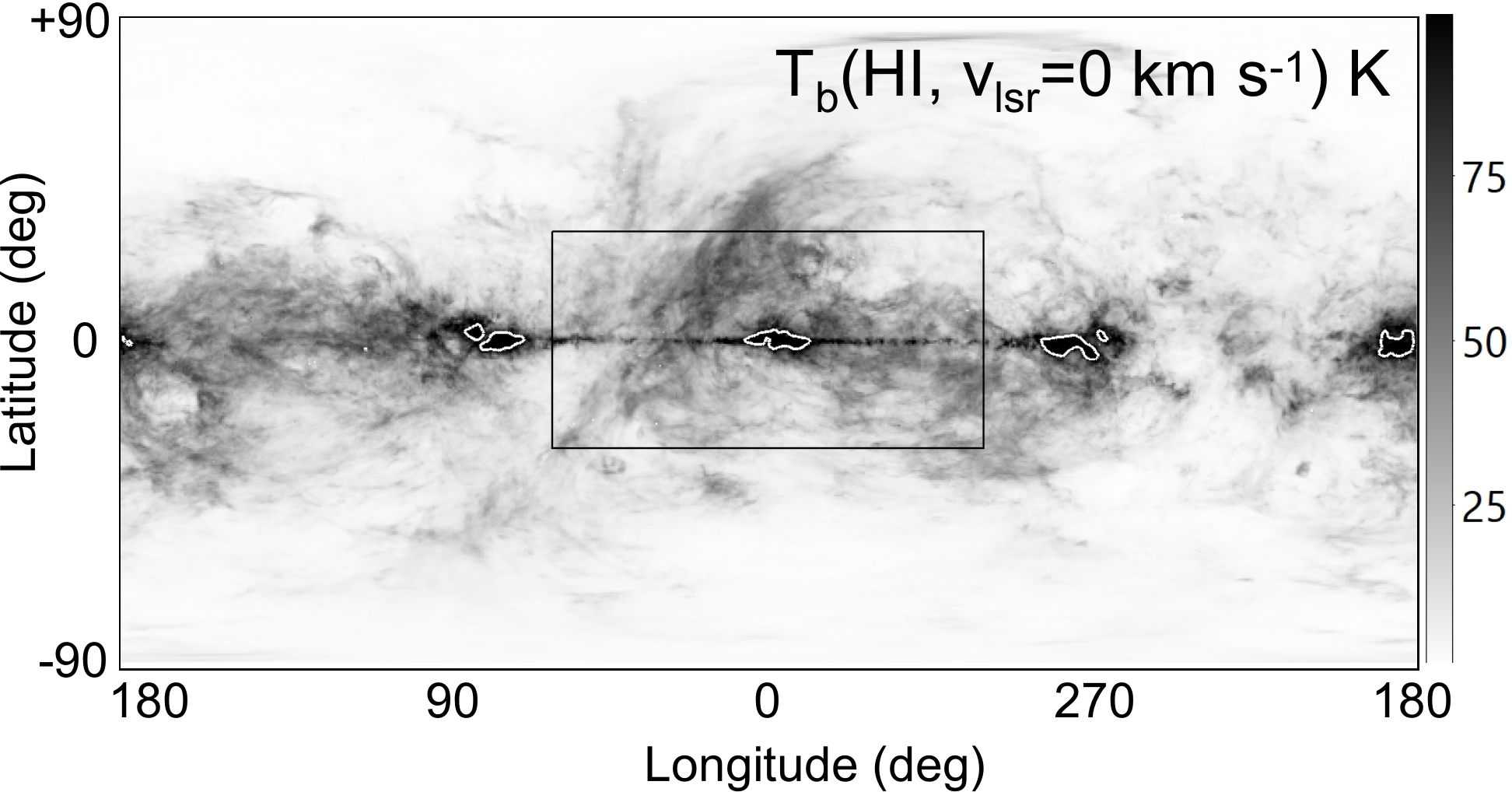}  
\end{center}
\caption{HI $\Tb$ distribution on the sky at $\vlsr=0$ \kms from the LAB HI survey (Kalberla et al. 2005). White contours are at $\Tb=100$ K, approximately enclosing the radial velocity-degenerate regions (VDR). The square shows the analyzed region in this study. }
\label{area} 
	\end{figure}

\begin{figure}
\begin{center}
\includegraphics[width=7cm]{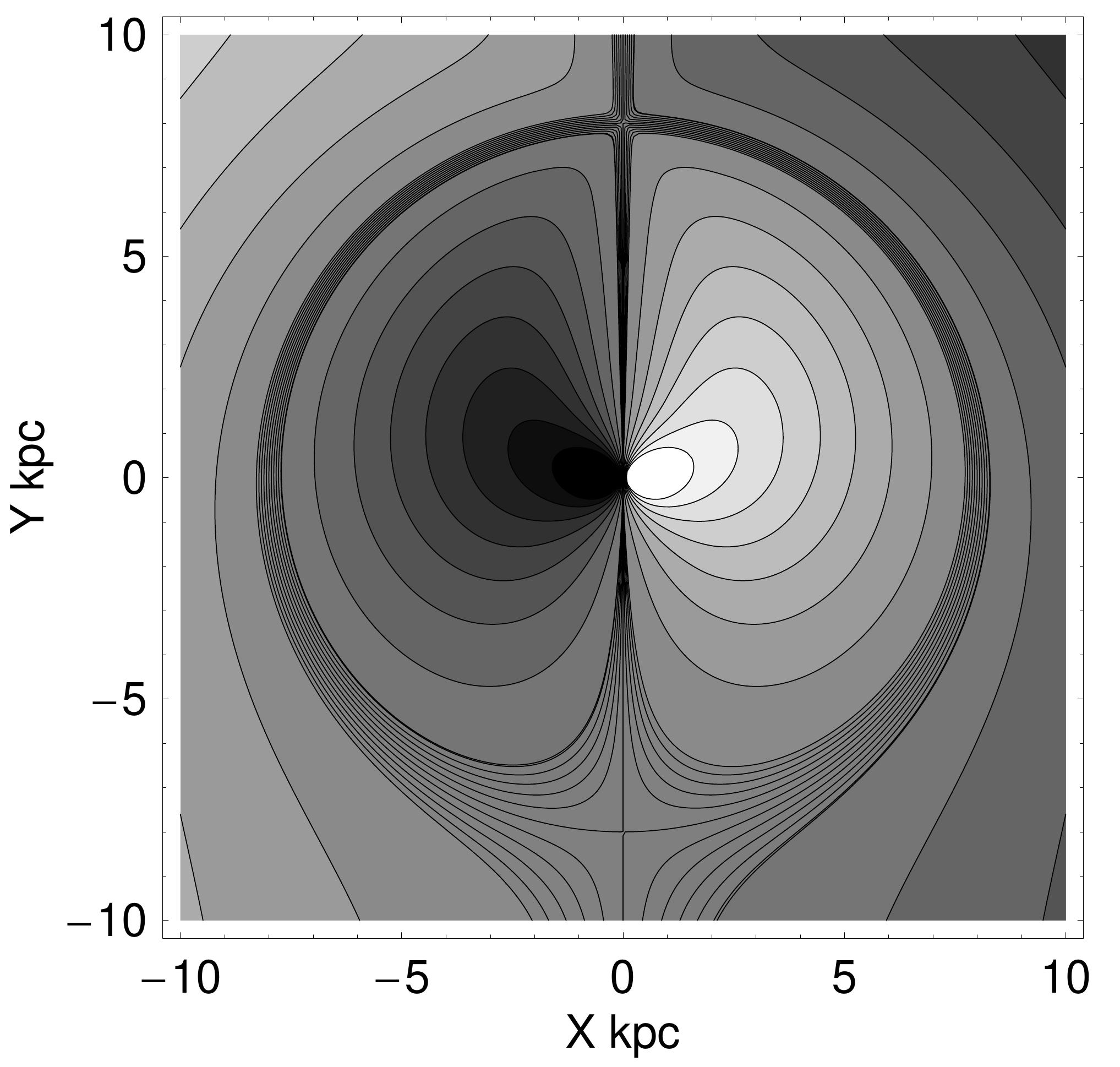}  
\end{center}
\caption{Velocity-degenerate region (VDR) at $-5\le \vlsr \le +5$ \kms indicated by dense contours with 1 \kms interval in the radial velocity field of the Galaxy calculated for the rotation curve of Miyamoto and Nagai's (1975) potential. Other contours are at every $\pm 20$ \kms. The Sun is at $X=0,\ Y=R_0=8$ kpc, and the rotation velocity is $V_0=200$ \kms. } 
\label{VDR}
	\end{figure}

Figure \ref{area} shows the distribution of HI line brightness temperature, $\Tb$, at $\vlsr=0$ \kms as taken from the LAB HI survey at spatial resolution of $0\deg.6$ FWHM (Kalberla et al. 2003). Since the channel velocity interval is 1.0 \kms and the velocity resolution of the survey was 1.3 \kms, the emission comes approximately from the shaded region shown in figure \ref{VDR}, considering the velocity dispersion of $\sim 10$ ($\pm 5$) \kms. Most of off-galactic plane emission comes from local HI gas near the Sun, whereas emission from the VDR, e.g. regions in the Galactic plane toward the GC, anti-GC and tangential directions of the solar circle, comes through long path in the Galactic disc (figure \ref{VDR}). In figure \ref{area} we show the VDR as regions approximately enclosed by contours at $\Tb=100$ K. We use the saturated HI brightness in the VDR to determine the spin temperature. 

For determination of the local HI gas density around the Sun, we analyze the rectangular region indicated in figure \ref{area}, and we use off-Galactic plane data at $|b|> 1\deg$ in order to avoid the contamination of far-side HI emission along the solar circle. 

The map partially includes the outskirt line emission from the Aquila Rift, which is located at $\sim 200$ pc away from the Sun corresponding to the center velocity of $\vlsr \sim 4$ \kms, having line width of several \kms (Sofue and Nakanishi 2017a). In order ease such contamination from individual objects, we will average the values in regions as wide as possible in the box shown in figure \ref{area}.

\subsection{Radio continuum map}

\begin{figure} 
\begin{center} 
\hskip-5mm\includegraphics[width=9cm]{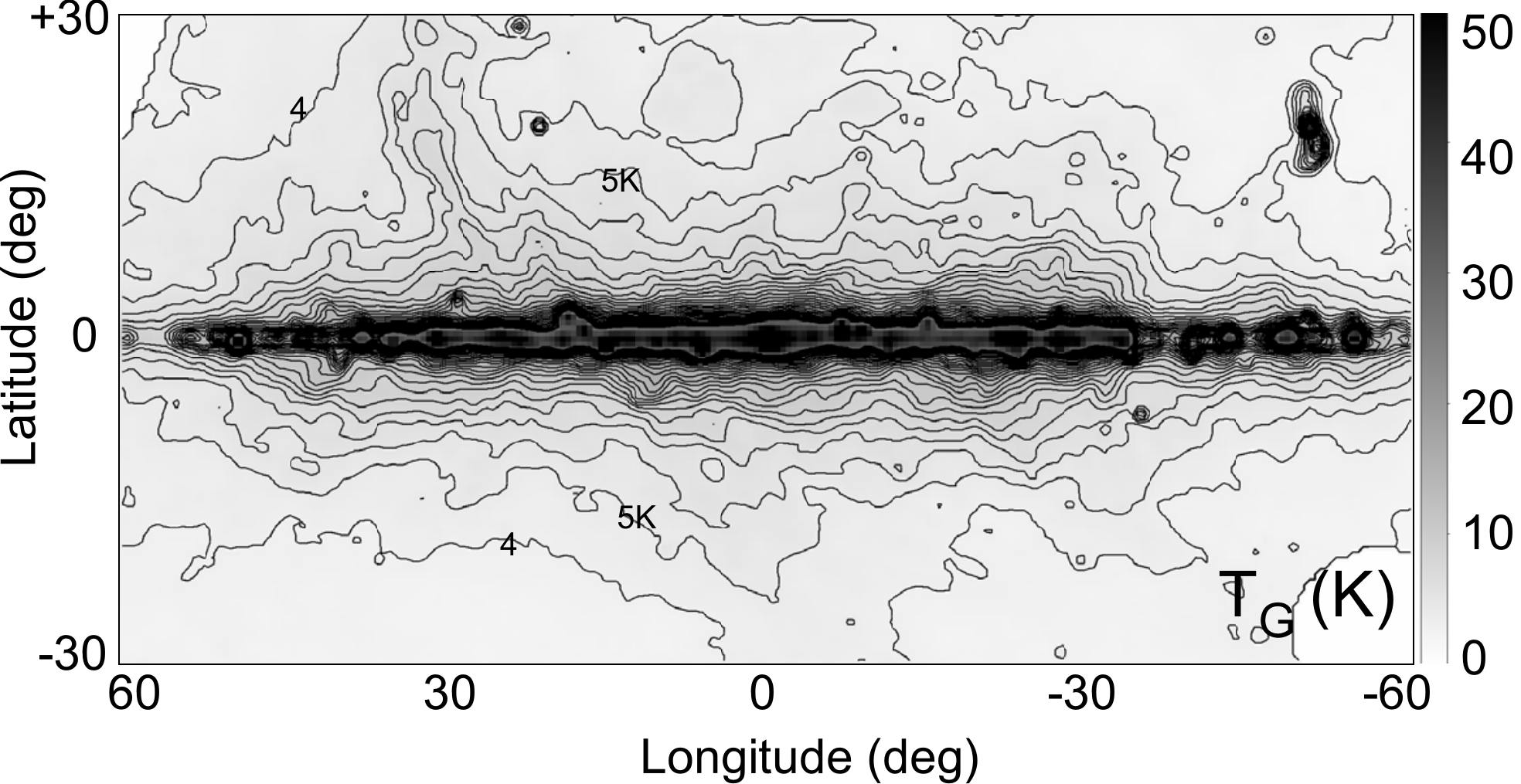} 
\end{center}
\caption{Galactic 1.4 GHz radio continuum map of $\Tg$, as converted from 2.3 GHz Rhodes map (Jonas et al. 1985) for an assumed spectral index $\beta=-2.7$. Contour interval is 1K. A map of $\Tc=\Tg+\Tcmb$ is used for analysis of local HI. }
\label{map-Tc} 
	\end{figure}

Figure \ref{map-Tc} shows the brightness distribution of radio continuum emission $\Tc$ at 1420 MHz, as converted from the Rhodes 2300 MHz Survey (Jonas et al. 1985). The brightness temperature at 2300 MHz was converted to that at 1420 MHz for an assumed spectral index of $\beta_{\rm synch}=-2.7$ as
\begin{equation}
\Tc=\Tc(\nu_1) (\nu_2/\nu_1)^{\beta_{\rm synch}}+\Tcmb,
\label{Tb1400}
\end{equation}
where $\nu_1=2300 MHz$ and $\nu_2=1420 MHz$. The assumption of synchrotron emission in off-galactic plane region is plausible, because thermal sources such as HII regions are highly concentrated toward the galactic plane in a thin disk of full thickness of $\sim 90$ pc, or $|b|<\sim 0\deg.3$. 

{ 
On the other hand, for the analysis of emission in the Galactic plane, we assume that the galactic continuum emission is an equal mixture of thermal and nonthermal emissions, as given by
\begin{equation}
\Tc=\Tc(\nu_1) [\eta(\nu_2/\nu_1)^{\beta_{\rm synch}}+(1-\eta)(\nu_2/\nu_1)^{\beta_{\rm therm}}] +\Tcmb,
\label{synchtherm}
\end{equation}
where 
the spectral indices for synchrotron and thermal radiations are assumed to be $\beta_{\rm synch}=-2.7$ and $\beta_{\rm therm}=-2$, respectively.
 Here, $\eta$ is the fractional ratio of the nonthermal and thermal emissions at $\nu_1$. Separation of the non-thermal and thermal contributions of the radio continuum emission in the Galactic plane showed that the ratio of the non-thermal to thermal brightness temperatures at 600 MHz is $\sim 2.5$ at $|l|<\sim 30\deg$ (Large et al. 1961). This ratio corresponds to $\eta \sim 0.5$ at $\nu_1=2.3$ GHz for the above spectral indices. So, we here adopt this value in equation (\ref{synchtherm}).
 }

The 2300 MHz map was used in order to avoid the absorption effect by HI in the 1420 MHz map. We confirmed that the converted intensities at 1420 MHz are consistent with that directly observed at 1420 MHz continuum at $l=15\deg-30\deg$ (Sofue and Reich 1979).  .

\section{Method}
\subsection{HI Emission and Absorption}  

The observed brightness temperature $\Tb$ of the HI line emission from an HI emitting source with spin (excitation) temperature $\Ts$ is given by
\begin{equation}
 \Tb = (\Ts-\Tc)(1-e^{-\tau}),
\label{eqTb}
\end{equation}
where $\tau$ and $\Tc$ are the optical depth and brightness temperature of background continuum emission.
The optical depth is, therefore, a function of the spin, brightness and continuum temperatures as
\begin{equation}
\tau=-{\rm ln}\left(1-{\Tb \over \Ts-\Tc}\right).
\label{eqtau}
\end{equation}

When the HI emission source is in front of the continuum source, $\Tc$ is written as 
\begin{equation}
\Tc=\Tg+\Tcmb,
\end{equation}
where $\Tg$ is the Galactic background brightness, and $\Tcmb=2.7$ K is the cosmic microwave background.  
This equation applies even in  {the} case that the continuum and HI emissions originate from the same region in so far as the region is optically thick against HI. However, if the HI source is optically thin in this case, $\Tg$ may be reduced by a factor of $\sim$  {4}, because the practical HI absorption takes place against the farther-side half of the continuum source by the nearer-side half of the HI source. 

For  {the} optically thin case, the HI brightness is approximated by
\begin{equation}
\Tb\simeq (\Ts-\Tc)\tau.
\label{eqThin}
\end{equation}
Furthermore, if the continuum background can be neglected, or $\Ts \gg \Tc$, this relation reduces to the often quoted relation, 
\begin{equation}
\tau \simeq \Tb/\Ts.
\end{equation}

On the other hand, for  {the} optically thick case, $\tau \gg 1$, we have
\begin{equation}
\Tb\simeq \Ts-\Tc.
\label{eqThick}
\end{equation}
This relation will be used to measure $\Ts$ toward velocity-degenerate directions, where the optical depth is sufficiently large. This approximation is valid  even if the continuum and HI sources are located in the same region, because the continuum emission from regions with $\tau >1$ is regarded to be the background.

\subsection{Correction factor $\Hfactor$ for HI density}

The column density of H atoms, $N$, on the line of sight is calculated by
\begin{equation} 
N=-\Xhi \Ts \int {\rm ln}\left(1-{\Tb \over \Ts-\Tc}\right)dv,
\label{eqNthick}
\end{equation} 
where $\Xhi=1.82\times 10^{18} {\rm cm^{-2} (K~km~s^{-1})^{-1} }$ is the conversion factor.  
The volume density $n$ is related to $\Tb$ as  
\begin{equation} 
n={dN\over dx}= -\Xhi \Ts  {\rm ln} 
\left(1-{\Tb \over \Ts - \Tc}\right)
 {dv \over dx}.
\label{nthick}
\end{equation} 

If the HI source is optically thin, and the continuum background is weak enough, e.g. if  $\tau \ll 1$ or $\Tb \ll \Ts$ and $\Tc \ll \Ts$, these equations reduce to the often used formulae,
\begin{equation}
N_{\rm thin} \simeq \Xhi \Ihi=\Xhi \int \Tb dv,
\label{eqNthin}
\end{equation} 
and  
\begin{equation}
 n_{\rm thin} \simeq \Xhi \Tb {dv \over dx}.
 \label{nthin}
\end{equation}

We introduce a correction factor $\Hfactor$ defined by the ratio of HI densities calculated using equations \ref{nthick} and \ref{nthin}:
\begin{equation}
\Hfactor={n \over n_{\rm thin}}={\Ts \over \Tb}\tau=-{\Ts\over \Tb} {\rm ln}\left(1-{\Tb \over \Ts-\Tc}\right).
\label{eqeta}
\end{equation}  
When $\Tb \ll \Ts$ and $\Tc \ll \Tb$, these equations reduce to $\Hfactor \simeq 1$ and $\tau \simeq \Tb/\Ts$.

\begin{figure}
\begin{center}
\includegraphics[width=7cm]{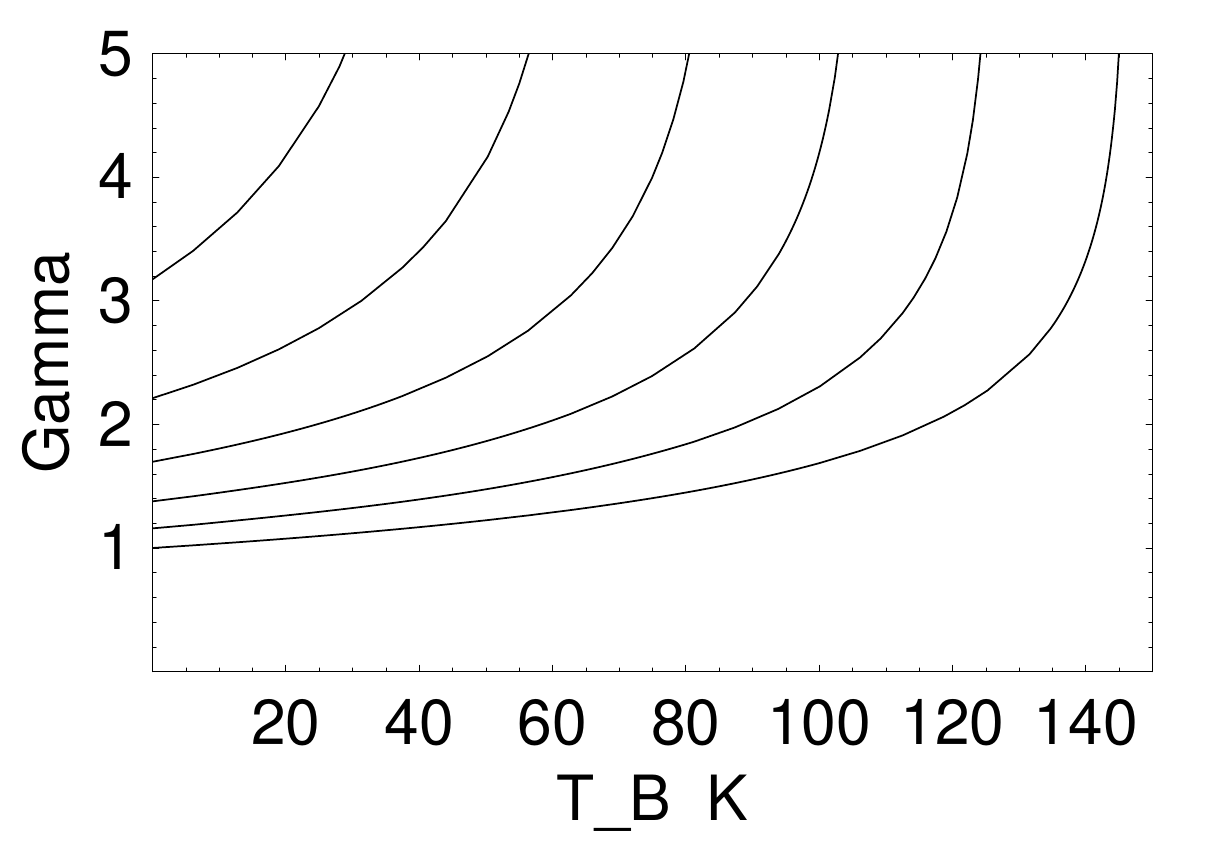} 
\end{center}
\caption{Correction factor $\Hfactor$ (ratio of volume HI density calculated for finite optical thickness to that for optically thin assumption) plotted against $\Tb$ for background continuum temperatures $\Tc=0$ (right bottom line), 20, 40, ... 100 K (top left) for spin temperature of $\Ts=146$ K.}
\label{ratio_theory}
	\end{figure}

The correction factor $\Hfactor$ may be used to obtain a more general and reliable HI density, when the HI line brightness temperature is not low enough compared to the spin temperature and/or the background continuum brightness is not negligible. Figure \ref{ratio_theory} shows the correction factor as a function of $\Tb$ calculated for several continuum brightness temperatures. 
 
\subsection{Local HI density}

Using the HI conversion factor $\Xhi$, we may express the volume density as
\begin{eqnarray}
n=5.89\times 10^{-4} \Tb{dv\over dx} \Hfactor , 
\label{eqVolDen}
\end{eqnarray}  
where $\Tb$ is measured in K and  ${dv/dx}$ in $ {\rm km\ s^{-1}kpc^{-1}}$.
When the gas is located in the solar vicinity, the velocity gradient is approximated by 
\begin{equation} 
{dv \over dx}\simeq A~ |{\rm sin~}2l| {\rm cos^2~} b + {d \sigma \over dx},
\end{equation}
where the first term represents the galactic rotation with { {$A=14.5\ {\rm km~s^{-1}~kpc^{-1}}$} being the Oort $A$ constant, and the line of sight derivative is related to the derivative in the galactic plane as $dx=dr/ \cos\ b$. The second term represents velocity gradient due to turbulent motion, and is approximated by
\begin{equation}
{d\sigma \over dx}\simeq {\sigma/\lambda},
\end{equation}
where $\lambda$ is the line of sight depth.

Using LAB HI line cube, we measured the full width of half maximum of the HI line profile at several positions in the analyzed region at $|b| >\sim 5 \deg$ to be $\sigma \sim 6$ \kms. The line of sight depth toward typical inter-VDR direction at $l\sim 45\deg$ would be $\lambda \sim \sigma/2A \sin 2l \sim 200\ $ pc corresponding to half $\sigma$ falling to the channel at $\vlsr=0$ \kms. We have thus $\sigma/\lambda \simeq 33$ \kms kpc$^{-1}$, and the volume density can be estimated by
\begin{equation} 
n\simeq 5.89\times 10^{-4} \Tb(14.5 |\sin\ 2l | \cos^2 b + 33)  \Hfactor,
\label{eqvol}
\end{equation}  
with $n$ measured in ${\rm cm^{-3}}$ and $\Tb$ in K.

\section{Spin Temperature}\label{secTs}

\subsection{Methods to determine $\Ts$}

In order to calculate the correction factor $\Hfactor$ we need to know the spin temperature $\Ts$. There have been a number of measurements of $\Ts$ in the ISM using emission and absorption spectra of the HI line against radio continuum sources. The measured temperatures for cold neutral ISM (CNM) range from 20 to 300 K, and those for warm neutral hydrogen (WNM) from $\sim 2000$ to $10^4$ K (see the literature in Section 1).  
 
The spin temperature of HI gas at a given frequency (velocity) can be measured by several methods. The simplest method is to compare the HI brightness temperature toward a cloud, $\Tb=(\Ts-\Tc)(1-e^{-\tau})$, with that toward an extragalactic or galactic continuum source located close enough inside the HI cloud, $\Tbon=(\Ts-\Tc-\Tcon)(1-e^{-\tau})$ 
 (Kuchar et al. 1991; Liszt et al. 1993; Heiles and Troland 2003a, b; Brown et al. 2014; Murray et al. 2014, 2015).
Here, $\Tcon$ represents the continuum brightness temperature toward the absorbing continuum source, which is measured at a nearby off-line frequency. 
 The continuum temperature, $\Tcon$, may be replaced by an absorbed brightness temperature, $\Taon$, by a foreground dense cloud as measured from the spectral self-absorption feature, where the HI density of the cloud has been estimated from molecular line observations ( {Li and Goldsmith 2003}; Goldsmith and Li 2005). 
 
Using these two relations with observed $\Tb$ and $\Tbon$, we can eliminate the term including $\tau$, and can obtain the spin temperature.  
 This method has the advantage that the measurement is accurate, if the continuum source is bright enough. On the other hand, its usage is limited only to HI clouds, where a radio continuum source is present close enough to the line of sight along which the measurement of $\Ts$ is made. 

An alternative method, which we propose in this paper, is the VDR method using saturated brightness temperature of the HI line in optically thick HI regions. For a sufficiently large optical depth ($\tau \gg 1$, equation (\ref{eqThick})), the spin temperature, $\Ts$, is related to the observed HI brightness temperature, $\Tb$, and background continuum brightness, $\Tc$, as
\begin{equation}
\Ts \simeq \Tb+\Tc.
\label{eq_vdr}
\end{equation}
This simple method has the advantage that it can be applied to widely distributed HI gas in the Galactic disk. On the other hand, it is restricted to regions with radial velocity ranges in the VDR, where the optical depth is sufficiently large. 

In the following, we apply this method to the VDR along the solar circle and the GC-Sun-anti Centre line using the LAB HI brightness distribution at $\vlsr=0$ \kms as shown in figure \ref{VDR}. Another advantage of this method is that we need no individual spectra, but we use only the brightness temperature $\Tb$ at a single frequency corresponding to the degenerate $\vlsr$.

\subsection{Saturated $\Tb$ in VDR}

Considering the velocity dispersion, the local HI emission at $\vlsr\sim 0$ \kms comes from the narrow region enclosed by equal-$\vlsr$ contours at a few \kms. The line of sight depth becomes large toward the four directions at $l \sim 0\deg, \ \pm 90\deg,$ and $\sim 180\deg$, which we call the velocity-degenerate region (VDR: figure \ref{VDR}).  

We now consider the VDR in the directions of the Galactic Center (GC VDR) and parallel directions to the solar circle (SC VDR). In such regions the column density at $\vlsr=0$ \kms is so high that the optical depth is large for CNM with the brightness temperature being nearly saturated at $\Tb \sim \Ts+\Tc$. The WNM is almost optically thin, because the absorption coefficient is inversely proportional to the spin temperature, which is as high as several thousand K.

In figure \ref{area} we show the distribution of HI brightness temperature, $\Tb$, at $\vlsr=0$ \kms on the sky from the LAB HI survey. The white contours are drawn at $\Tb=100$ K, which approximately enclose the four VDR. Figure \ref{gpTb} shows $\Tb$ distribution along the galactic plane, exhibiting saturated brightness in the VDR.
 
The line-of-sight depths in the VDR are of the order of $\sim 5$ kpc in the solar-circle, $20$ kpc in the GC. The column density along such depths for currently known HI density of $n\sim 1$ H cm$^{-3}$ is estimated to be on the order of $N\sim  (2 - 5)\times 10^{22}$ H cm$^{-2}$.
If the HI is CNM, the optical depth along such a long sight line and column is much greater than unity, and the measured maximum $\Tb$ can be regarded to represent the spin temperature.

On the other hand, if the gas is WNM, the brightness temperature must be as high as $\Tb \sim 10^3$ K for such large column density, which is not the case in the four VDR, even in the GC VDR. 
We may thus conclude that the assumption of CNM is plausible for the HI gas at $\vlsr \sim 0$ \kms, and the saturated values of $\Tb+\Tc$ in the VDR can be regarded to represent the spin temperature.

We, then, measured the saturated values of $\Tb+\Tc$ in the VDR in elliptical regions in the GC and local arm directions with $l$ and $b$ widths as listed in table \ref{tabTx}. The regions approximately trace the peak contours in figure \ref{area}. Thereby, we assumed that $\Tg$ is a mixture of thermal and nonthermal emissions as given by equation (\ref{synchtherm}).

By assuming that the saturated brightness represents the spin temperature, we thus obtained $\Ts=146.2\pm 16.1$ K and $144.4\pm 6.8$ K in the VDR toward the GC and local arm direction, respectively. The measured values are found to be nearly constant around 145 K. 

	\begin{table}
\begin{center}
\caption{Spin temperature, $\Ts$, measured in the VDR. } 
\label{tabTx}
\begin{tabular}{lll}  
\hline  
\hline
Method / Direction   & $\Ts$ (K) & $n$ (H cm$^{-3}$)\\ 
\hline   
Saturated $\Tb+\Tc$ in VDR$^\ddagger$ \\
GC ($10\deg \times 1\deg$) & $146.2 \pm 16.1$ & ---\\ 
Local Arm$^\dagger$, $l=277\deg$ ($3\deg \times 1\deg$) & $144.4 \pm 6.8$ & ---\\    
\hline  
 {$\chi^2$} fitting in the \\
Gal. Plane, $-30\deg \le l \le +30\deg$ 
 & $146.8 \pm 10.7$ & $0.89\pm 0.14$ \\ 
\hline 
\end{tabular}
\label{tabTs}
\end{center}   

\noindent $^\ddagger$ Average in narrow boxes of ($\Delta l \times \Delta b)$. \\
\noindent $^\dagger$ Rhodes continuum data not available in the Cygnus arm.
	\end{table}
 
\subsection{Model $\Tb$ distribution}

Given the spin temperature and hydrogen density $n$, the HI brightness temperature at $\vlsr=0$ \kms along the galactic plane is expressed by
\begin{equation}
\Tb=(\Ts-\Tc) (1-e^{-n\ r_v/\Xhi \Delta v\ \Ts}),
\label{Tb_calc}
\end{equation}
Here, $r_v$ is the line-of-sight depth of the region enclosed by equal-velocity contours in the velocity field, and is given by
\begin{equation}
r_v=|\Delta v/(A \sin 2l)|.
\label{rv}
\end{equation}
Thus, we have
\begin{equation}
\Tb=(\Ts-\Tc) [1-e^{-n\ /(A\ \sin2l\ \Xhi\ \Ts)}].
\label{Tb_calcA}
\end{equation}
This equation gives a good approximation in the GC direction, but it over-estimates the brightness at $l\sim \pm 90\deg$, where the line-of-sight depths are finite and the deepest sight lines point to $l\sim \pm 80\deg$.

The continuum contribution in the Galactic plane can be expressed as
\begin{equation}
\Tc\simeq 16\  e^{-(l/50\deg)^2}  +2.7 \ ({\rm K}),
\label{eqmodel}
\end{equation}
approximately fitting to the observed $\Tc$ in the case of mixture of thermal and nonthermal emission. 

We now calculate model $\Tb$ distribution as a function of longitude for an assumed density of $n=0.8$ H cm$^{-3}$.
In figure \ref{gpTb} we show the calculated model curves for several $\Ts$ from 50 to 1000 K, and for optically thin case. In the lower panel we enlarge the plots for the VDR for $\Ts=100$ to 160 K every 10 K. The model curve for $\Ts=140-150$ K well reproduces the general characteristics of the observed $\Tb$ distribution.. 

According to the approximation by equation (\ref{rv}), the model $\Tb$ profile is identical in the four directions at $l=0\deg,\ 90\deg,\ 180\deg$ and $270\deg$ except for the slight decrease in the GC direction due to stronger continuum emission. This approximation overestimates the line of sight depth in the solar-circle directions, and the observed peaks at $l\pm 75\deg$ are much lower than the model values. In the anti-center VDR ($l\sim 180\deg$) the observed $\Tb$ is also lower, which is due to lower density in the outer Galaxy.

 Model curves for higher spin temperature than $\ge \sim 150$ K, including high temperatures corresponding to WNM, cannot reproduce the observation. Also, curves for lower temperature than $\le \sim 120$ K cannot fit the observation. We may thus confirm that the estimated local spin temperature of $\Ts \sim 145$ K is  reasonable.

	\begin{figure} 
\begin{center}  
\includegraphics[width=9cm]{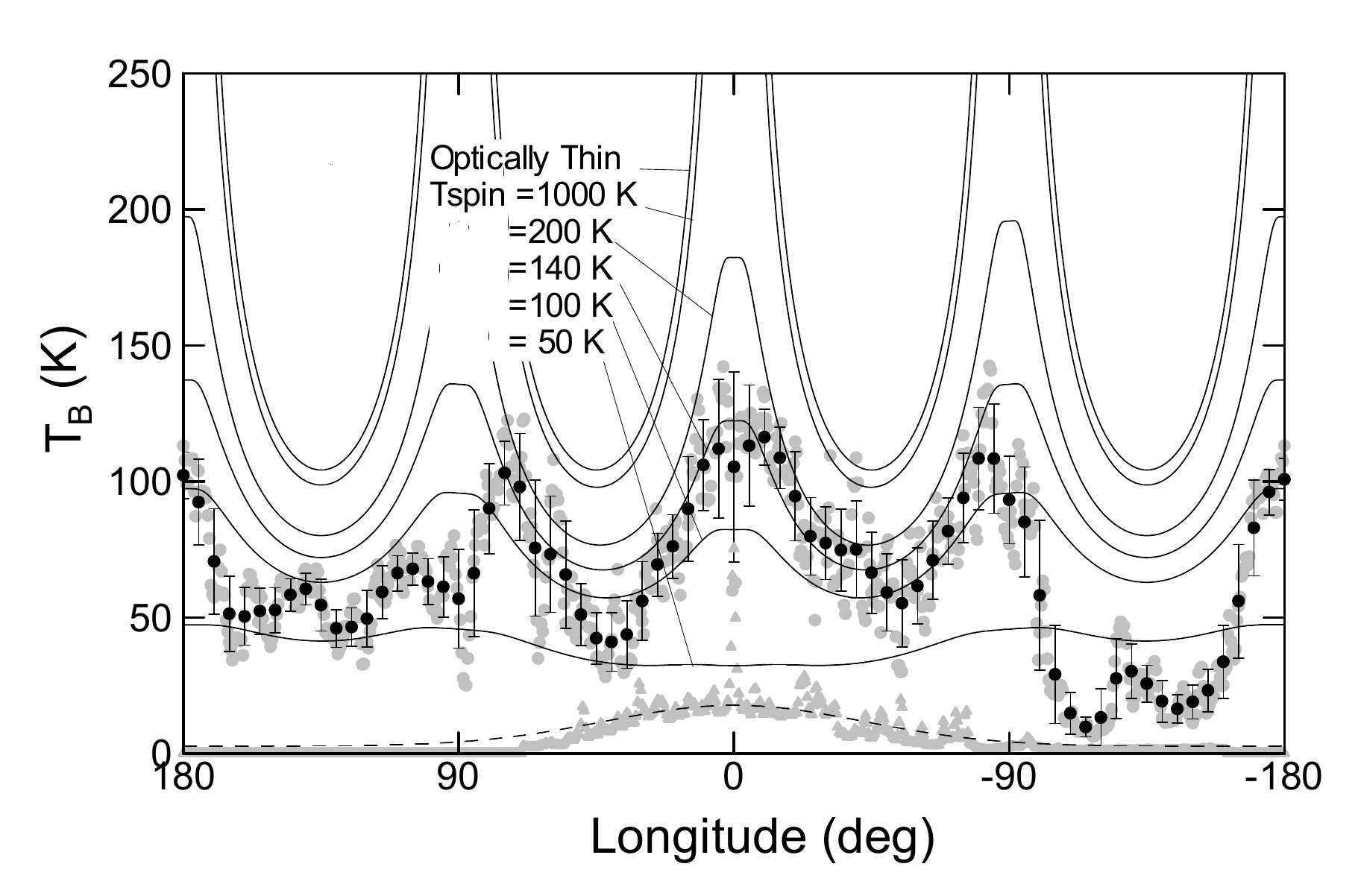}      
\includegraphics[width=9cm]{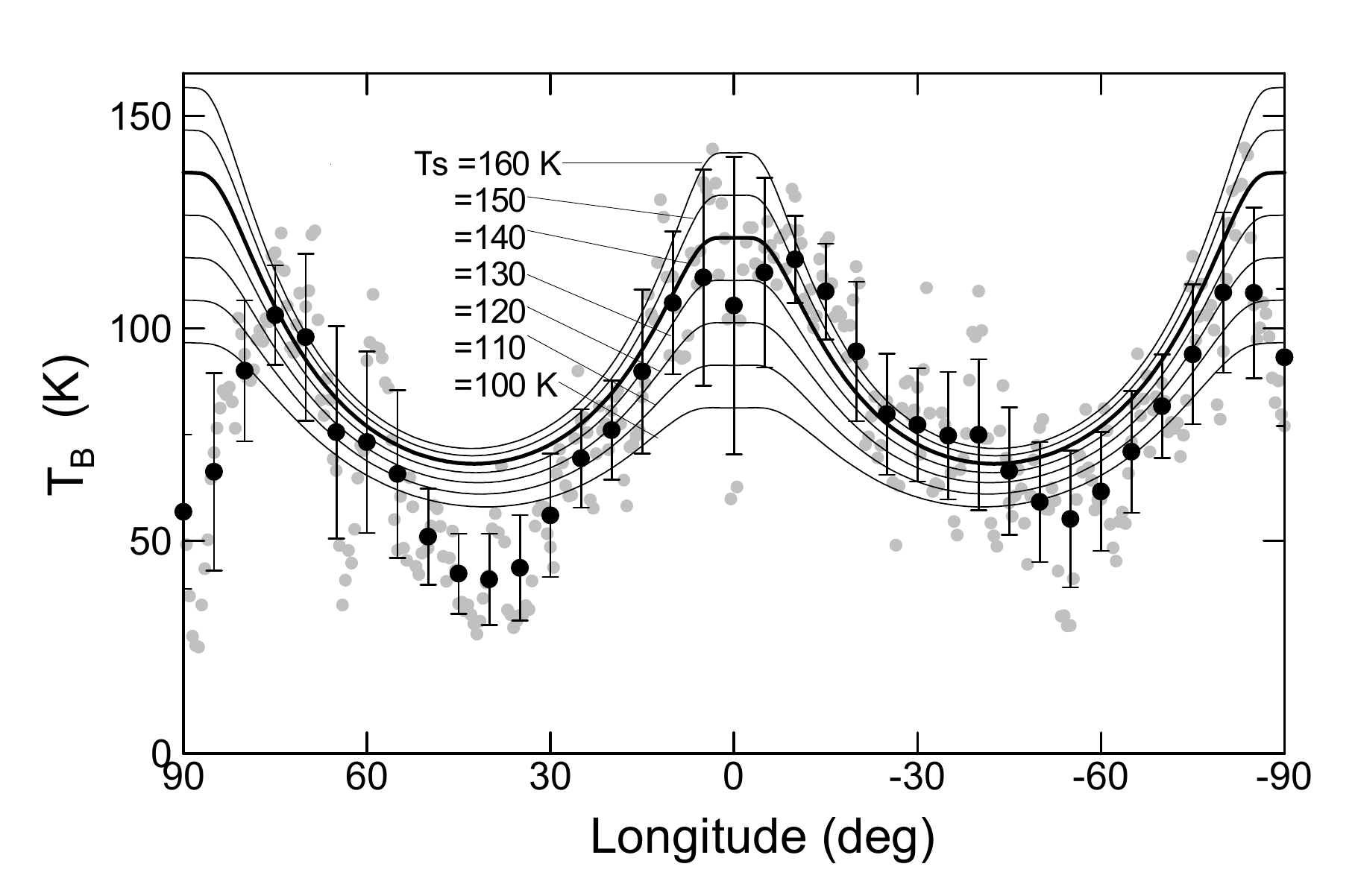}  
\end{center}
\caption{[Top] $\Tb$ along the Galactic Plane at $\vlsr=0$ \kms for fixed HI density $n=0.8$ H cm$^{-3}$ and velocity range $\Delta v=5$ \kms calculated for different $\Ts$ from 60 to 1000 K (smooth curves). Black circles show Gaussian-smoothed observed $\Tb$ at $5\deg$ longitude interval. Bars are standard deviations. Grey dots show raw $\Tb$ before smoothing. Continuum brightness $\Tc$ is shown by grey triangles, which was approximated by the thin dashed curve in the calculation.
[Bottom] Same, but enlarged for observed and model $\Tb$ calculated for $\Ts=100$ to 160 K every 10 K.  The model curve for $\Ts=130$ K (thick curve) fits the observations in the VDR well.}
\label{gpTb}
	\end{figure}

\subsection{$\chi^2$ fit to $\Tb$ distribution}

We now try to determine the spin temperature in a more statistically way by fitting the $\Tb$ distribution by least $\chi^2$ method. Thereby, the spin temperature $\Ts$ and the HI volume density $n$ are taken as the two free parameters. 
Using equation (\ref{Tb_calcA}), we calculate the $\chi^2$ by
\begin{equation}
 \chi^2 =\Sigma_i ({\Tb}_{,i}-{\Tb}_{\rm , calc})^2/\sigma_i^2.
 \end{equation}
Here, ${\Tb}_{,i}$ and $\sigma_i$ are the measured brightness temperature and its standard deviation as plotted in figure \ref{gpTb} by black dots and error bars, and ${\Tb}_{\rm , calc}$ is calculated temperature using equation (\ref{Tb_calcA}) that includes $\Ts$ and $n$ as the two free parameters. 
We, thus, apply the $\chi^2$ fitting to data at $-30\deg \le l \le +30\deg$, where equation (\ref{rv}) gives good approximation. Here, we used observed $\Tc$ instead of the model by equation (\ref{eqmodel}).

In figure \ref{ChiSqr} we show the calculated variation of $\chi^2$ in the $\Ts -n$ space. The best fit pair of $\Ts$ and $n$ was so determined that it made the least $\chi^2$ minimum. The errors are the deviations from the best-fit values that increase the $\chi^2$ by 1 from its minimum. 

We thus obtain a spin temperature of $\Ts=146.8 \pm 10.7$ K and hydrogen density $n=0.89\pm 0.14$. We list the values in table \ref{tabTs} in comparison with the other estimations using the saturated temperatures in the VDR. A simple mean of the listed three values with an equal weight yields $\Ts=145.8 \pm 1.2$ K, and we adopt this temperature for the analyses of $\Hfactor$ and $\tau$.  
 
	\begin{figure} 
\begin{center}     
\includegraphics[width=7cm]{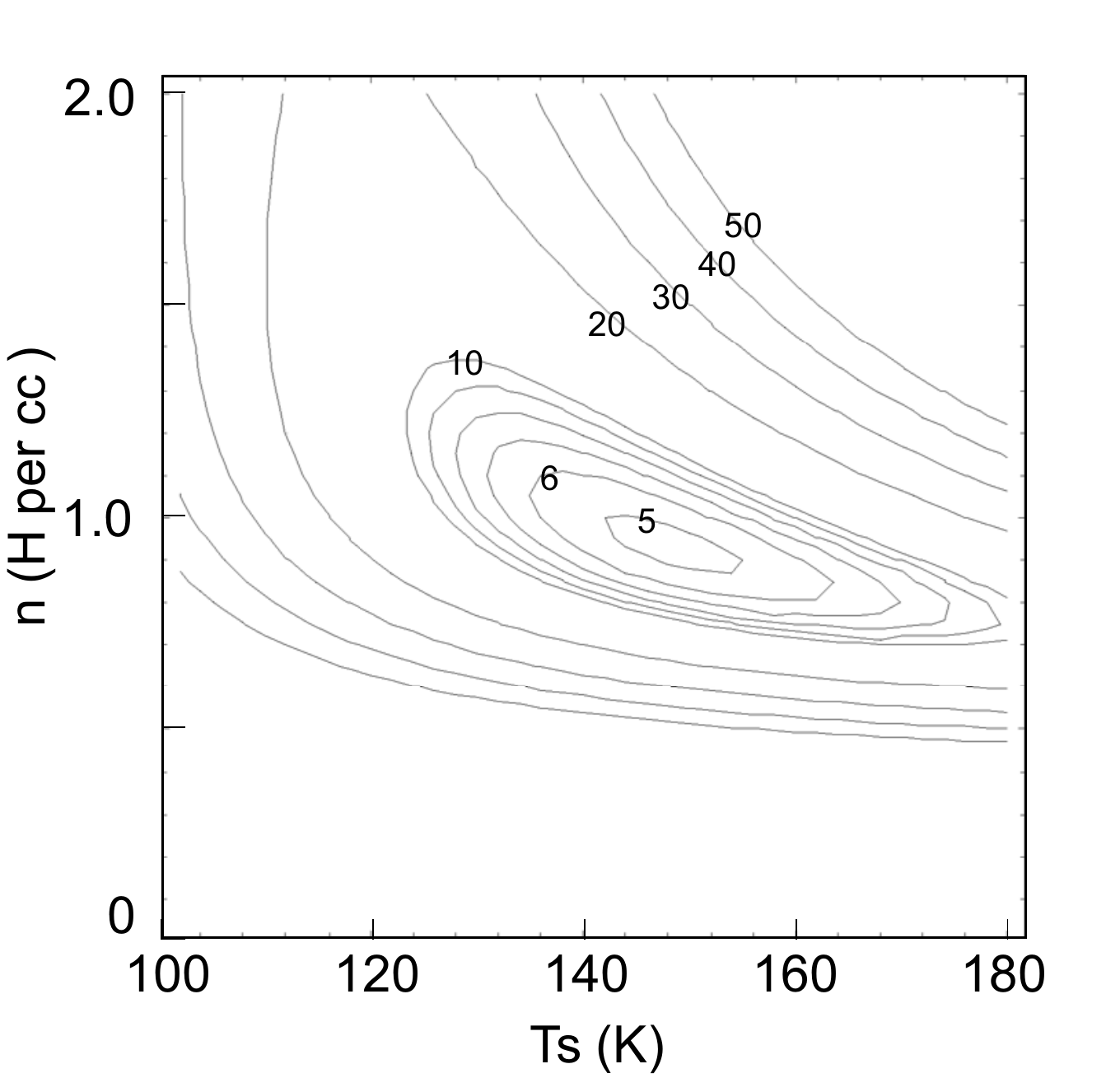}   
\end{center}
\caption{Contour diagram of $\chi^2$ in the ($\Ts, n$) plane. The best-fit set of the parameters yields $\Ts=146.8\pm 10.7$ and $n=0.89\pm 0.14$ at the $\chi^2$ minimum (=4.75). The errors are the deviations of the parameters from the best-fit values that increase $\chi^2$ by 1.  }
\label{ChiSqr}
	\end{figure}

\section{Off-Plane Analysis for Local HI}

\subsection{$\Hfactor$, $\tau$, and corrected local HI map} 

Using the observed HI and continuum brightness temperatures, we calculated the distribution of the optical depth $\tau$ and the correction factor $\Hfactor$ on the sky as shown in figures \ref{map-Tau} and \ref{map-Eta}, respectively. 

	\begin{figure}
\begin{center}
\hskip-5mm\includegraphics[width=9cm]{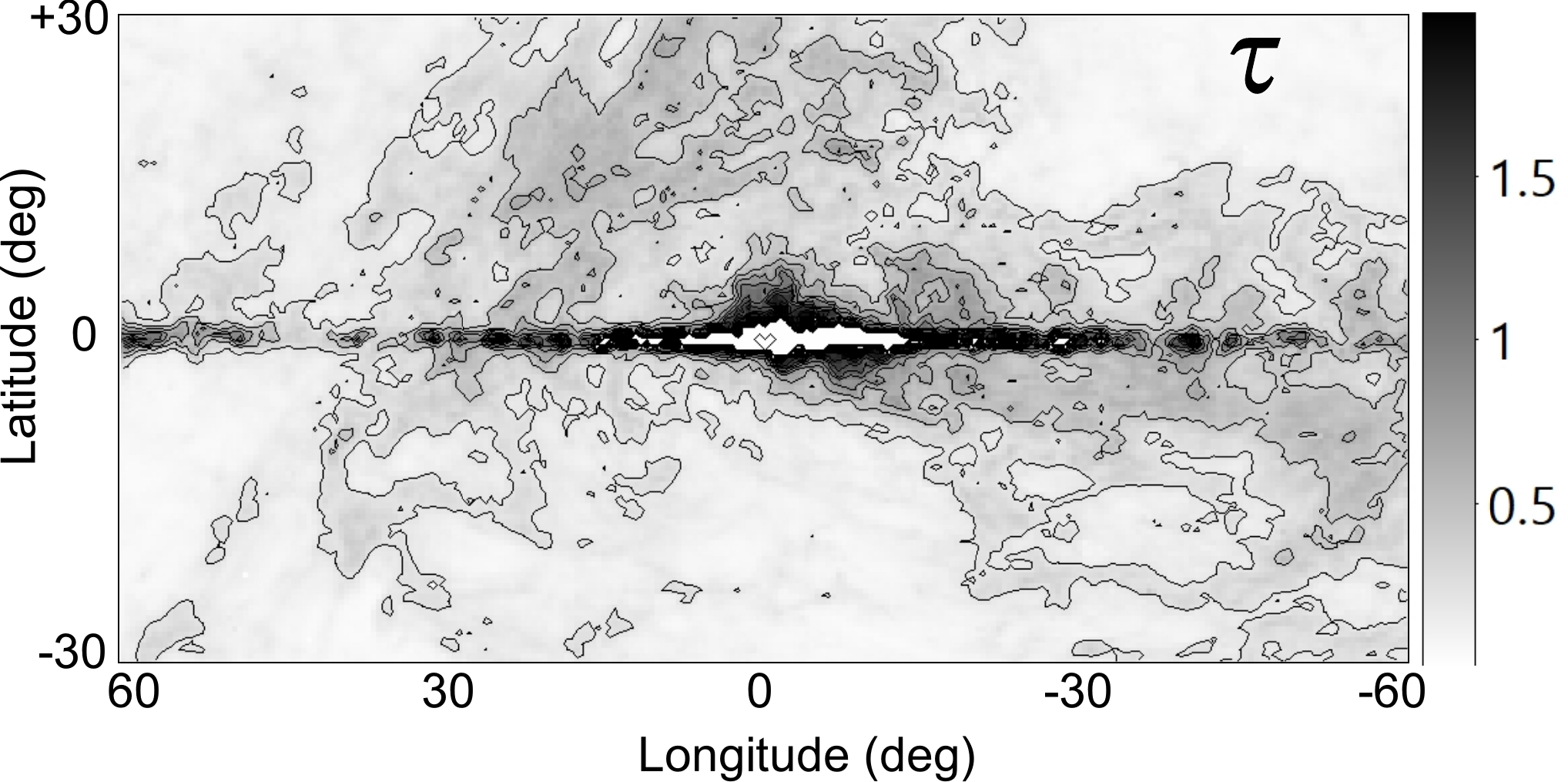}  
\end{center}
\caption{Optical depth $\tau$. Contours start from 0.2 at interval 0.2.}
\label{map-Tau}

\begin{center}
\hskip-5mm\includegraphics[width=9cm]{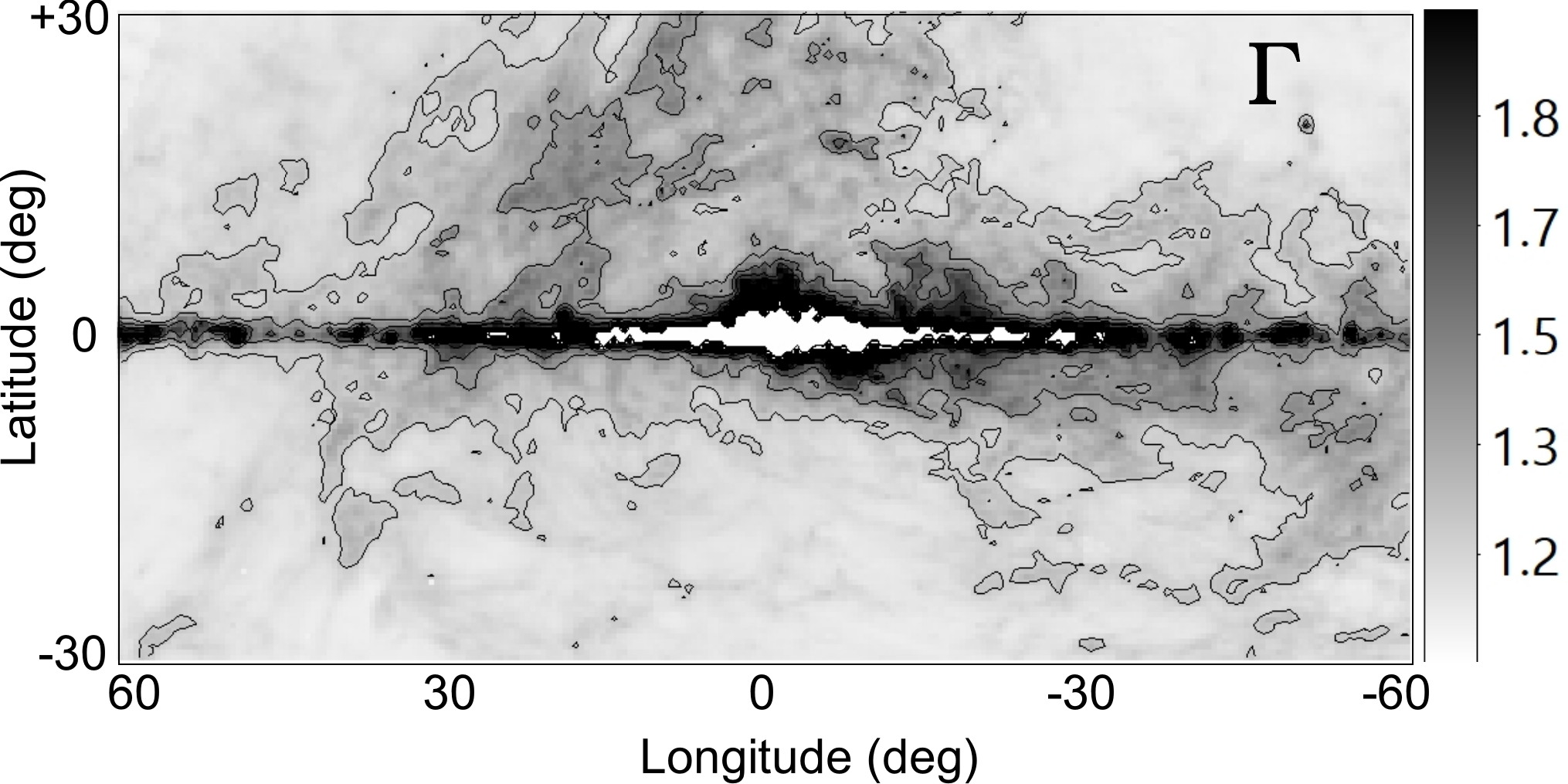} 
\end{center}
\caption{Grey-scale map of the correction factor $\Hfactor$. Contours start from 1.2 at interval 0.2.}
\label{map-Eta} 
	\end{figure}   
        
  Figure \ref{Tbhicon} shows variation of $\Hfactor$ against longitude at fixed latitude $b=2\deg.5$ compared to those for the brightness temperatures of HI and continuum emissions. Figure \ref{eta-b} shows $\Hfactor$ distribution at different latitudes.        
   
 The $\Hfactor$ value is observed to be around $\Hfactor \sim 1.2$ at $b>\sim 2\deg.5$ and $|l|>\sim 30\deg$. It gets larger near the galactic plane at $b<2\deg.5$.
In some regions toward known HI clouds, $\Hfactor$ attains higher values even at higher latitudes. 

In table \ref{tabAv} we list typical values averaged in the off-plane regions enclosed by the box shown in figure \ref{area}. An averaged value of $\Gamma \sim 1.2$ was obtained for the local HI gas. The slightly larger value in the northern half of the region would be due to the contamination of the Aquila Rift.
  
	\begin{figure} 
\begin{center} 
\includegraphics[width=8cm]{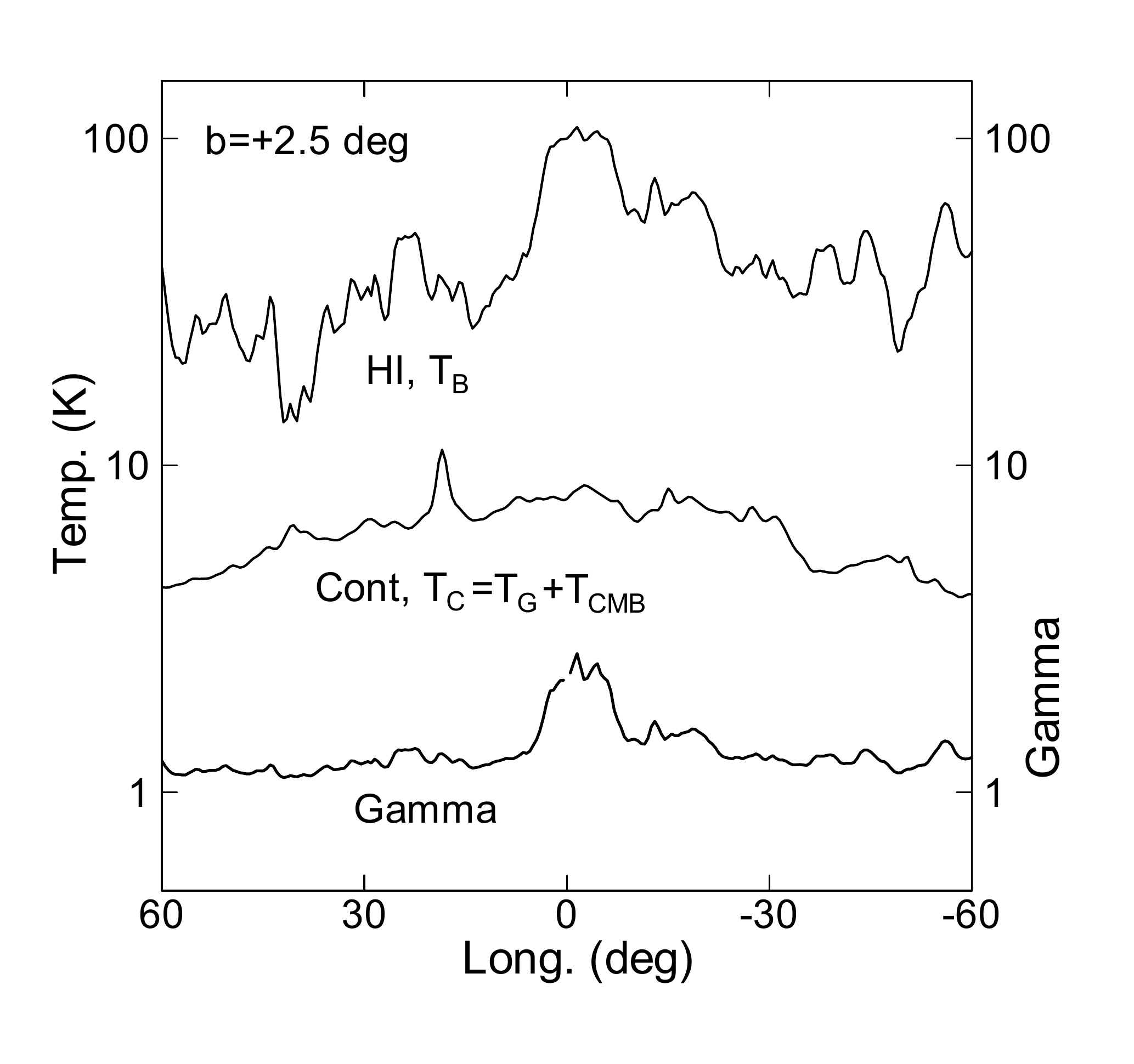}  
\end{center}
\caption{HI $\Tb$ at $\vlsr = 0$ \kms, $\Tc$ at 1420 MHz, and the correction factor $\Hfactor$, plotted against $l$ at $b=+2.5\deg$. }
\label{Tbhicon}
        
\begin{center}
\includegraphics[width=8cm]{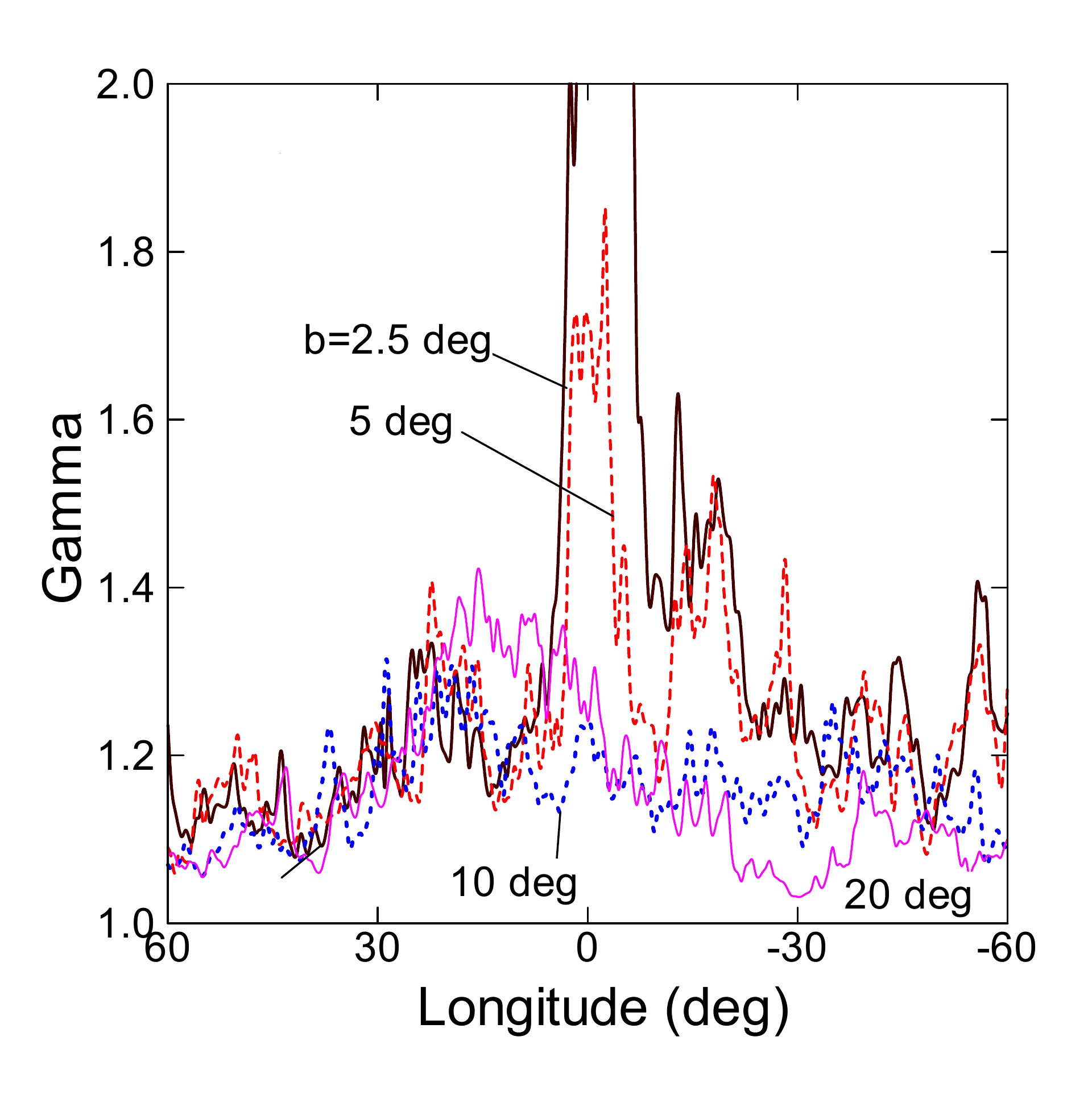}     
\end{center}
\caption{$\Hfactor$ against $l$ at different $b$. } 
\label{eta-b} 
	\end{figure}

	\begin{table}
\begin{center}
\caption{Mean and standard deviation of the correction factor $\Hfactor$ and optical depth $\tau$.}
\label{tabAv} 
\begin{tabular}{lll}  
\hline 
\hline 
Region, $l\deg\pm \Delta l \ / \ b\deg$ range & $\Hfactor$ & $\tau$ \\
\hline 
\hline
OFF-Galactic Plane  \\ 
~~~ $0.0 \pm 60 \ / \ b=-30\sim -5$&$1.22\pm 0.12$ &$0.24\pm 0.15$\\ 
~~~ $0.0 \pm 60 \ / \ b=+5\sim +30$&$1.17\pm 0.08$ &$0.18\pm 0.11$\\ 
Gal. Plane Minima, Inter-VDR\\
~~~ $   40.0 \pm  10.0 $ & $   1.36 \pm   0.14 $ & $   0.43 \pm   0.18 $ \\
~~~ $  130.0 \pm  10.0 $ & $   1.26 \pm   0.06 $ & $   0.44 \pm   0.10 $ \\
~~~ $  230.0 \pm  10.0 $ & $   1.14 \pm   0.06 $ & $   0.20 \pm   0.10 $ \\
~~~ $  310.0 \pm  10.0 $ & $   1.47 \pm   0.17 $ & $   0.64 \pm   0.25 $ \\
\hline
Av.$^\dagger$ Local & $1.22 \pm 0.03$ & $ 0.30 \pm 0.05$ \\
\hline  
\hline
VDR, Sol. Circle (Arm)\\ 
~~~ $   75.0 \pm   5.0 $ & $   1.84 \pm   0.23 $ & $   1.32 \pm   0.31 $ \\   
~~~ $  280.0 \pm   5.0 $ & $   2.19 \pm   0.67 $ & $   1.68 \pm   0.81 $ \\ 
VDR, Anti-GC\\ 
~~~ $  180.0 \pm   5.0 $ & $   1.77 \pm   0.13 $ & $   1.23 \pm   0.18 $ \\ 
\hline
Av. Arm + Anti-GC & $1.80 \pm 0.11$ & $1.27 \pm 0.15$ \\ 
\hline
\hline
VDR,  GC\\
~~~ $    0.0 \pm   5.0 $ & $   3.63 \pm   0.82 $ & $   2.66 \pm   0.86 $ \\
~~~ $    0.0 \pm  10.0 $ & $   3.36 \pm   0.86 $ & $   2.53 \pm   0.86 $ \\
\hline
\end{tabular}\\
$^\dagger$ Average of the individual regions in the distance category using their standard deviations as the weights.
\end{center} 
	\end{table}   
 
Using the maps thus obtained for $\tau$ and $\Hfactor$ combined with the HI brightness map at $\vlsr$, we finally calculated the volume density of HI gas using equation (\ref{eqvol}). The result of the volume density distribution on the sky is shown in figure \ref{nHmap}.  

Using the map, avoiding the near-Galactic plane region, we determined the local HI density to be $n=0.86\pm 0.52$ H cm$^{-3}$ in the northern area at $b \ge 5\deg$, and $n=0.69\pm 0.46$  H cm$^{-3}$ in the south at $b\le 5\deg$.
These values are consistent with the value, $n=0.89\pm 0.14$ H cm$^{^3}$, as obtained by the $\chi^2$ fitting of $\Tb$ to the region around the VDR near the Galactic plane as shown in table 1.

	\begin{figure} 
\begin{center}
\hskip-5mm\includegraphics[width=9cm]{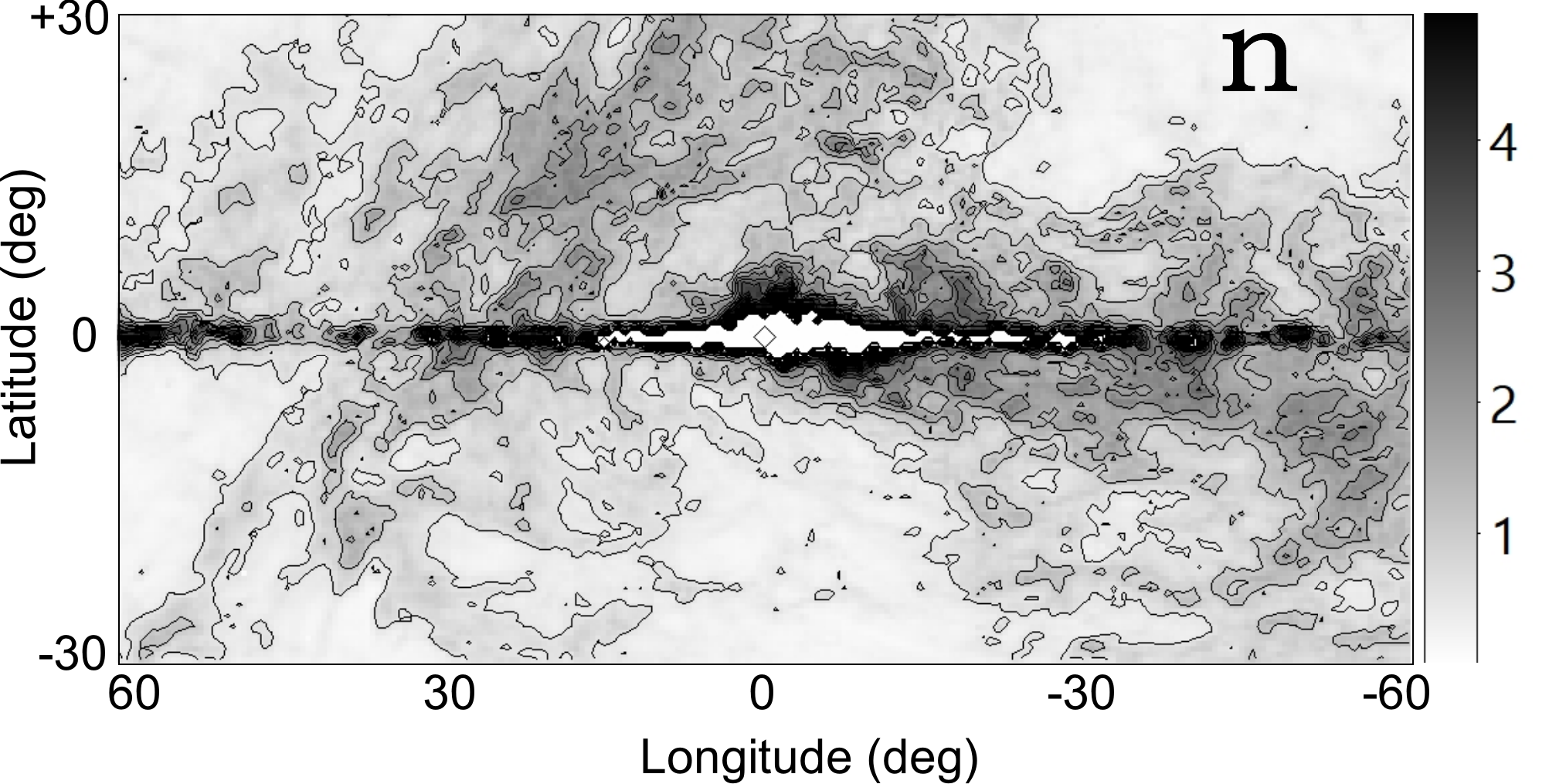}  
\end{center}
\caption{Distribution of the volume density of local HI gas, $n$. Contour interval is 0.5 H cm$^{-3}$.}
\label{nHmap}
\label{map-volden}
	\end{figure}

\section{Galactic-Plane Analysis for Global HI}

\subsection{$\Hfactor$ and $\tau$ in the Galactic Plane}

In the previous analysis  {of} local HI gas, we avoided the saturated regions in the VDR and the Galactic plane. However, equation (\ref{eqeta}) applies to any regions for measured $\Tb$ and $\Tc$. In fact figure \ref{eta-b} shows higher $\Hfactor$ toward the Galactic plane. 

In order to analyze $\Hfactor$ and $\tau$ in the Galactic plane, we replace the radio continuum brightness given by equation (\ref{Tb1400}) with brightness for a mixed case  {in which} thermal and synchrotron emissions coexist using equation (\ref{synchtherm}). In fact, thermal radio sources such as HII regions are tightly concentrated to a thin galactic disk of full thickness $\sim 90$ pc (Hou and Han 2014; Sofue and Nakanishi 2017b).

Note, however,  we put the galactic continuum emission to be zero in the longitude range from $l=70\deg$ to $200\deg$, where no Rhodes continuum data are available. According to equation (\ref{eqeta}), this approximation may cause a slight systematic under estimation of $\Gamma$ of the order of $\sim 10^{-2}$, but the amount is within the measurement error. This is because $\Tb$ in this longitude range is considered to be less than a few K and is much lower than $\Ts$.

We show the variation of the thus calculated $\Hfactor$ and $\tau$ along the Galactic plane in figure \ref{etaVDR}, where the values are averaged in longitude bins of $5\deg$ width. The bars are standard deviation in the bins. Grey dots show raw values before averaging.

	\begin{figure} 
\begin{center}  
\hskip -5mm\includegraphics[width=9cm]{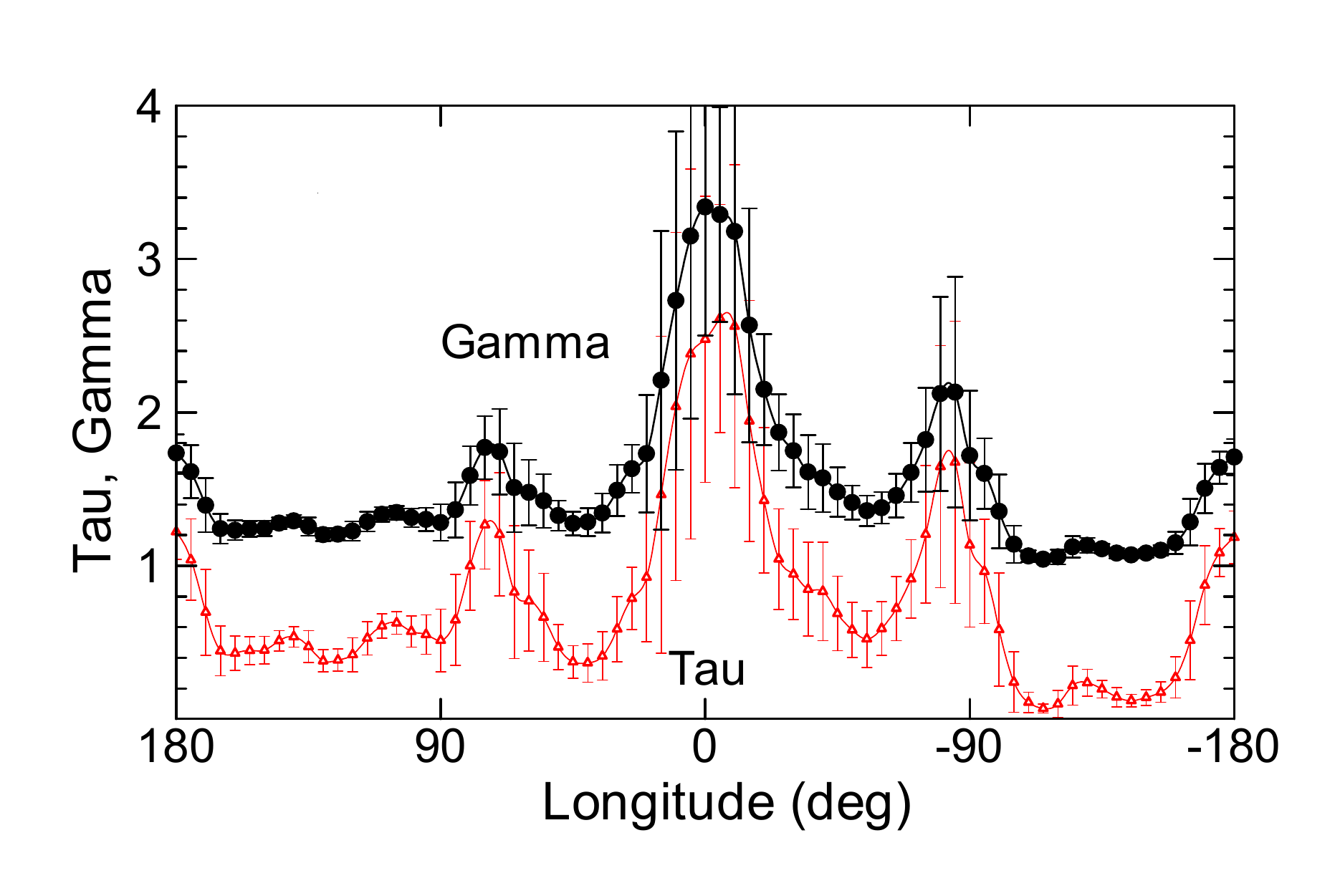}     
\end{center}
\caption{$\Hfactor$ (black circles) and $\tau$ (triangles) in the Galactic plane averaged in longitude bins of $\pm 5\deg$ width. Bars indicate standard deviations. }  
\label{etaVDR} 
	\end{figure}  

In table \ref{tabAv} we list the measured values of $\Hfactor$ and $\tau$ along the Galactic plane ($b=0\deg$). The listed values are mean $\Hfactor$ and $\tau$ in representative directions in the Galactic Plane ($b=0\deg)$ at $l$ within $\pm \Delta l$ from the indicated longitudes. The mean was obtained by taking Gaussian-running average in each longitudinal bin of half-width of $5\deg$. 

\subsection{Local value of $\Hfactor$ in the inter-VDR directions}

Values close to unity with $\Hfactor \sim 1.2$ was obtained in the off-arm directions, where the line of sight depth is less than a few hundred pc corresponding to the velocity range around $\vlsr=0$ \kms. Therefore, the local HI gas is optically thin even in the Galactic plane, and is consistent with the off-plane local values as listed in table \ref{tabAv}.
The values obtained in individual regions, including those for the off-galactic plane regions, are further averaged using their standard deviations as weights, and listed as the mean of the local $\Gamma$ and $\tau$ values in the row below the category in table \ref{tabAv}.  

\subsection{Modest $\Hfactor$ in the local arm direction}

Values about $\Hfactor\sim 2$ were obtained in the directions of the local arms as well as in the anti-center direction. The line of sight depths in these directions are several kpc, and the HI mass in the disk in the vicinity of the Sun may be doubled. This value may be compared with the values observed for HI gas in circum molecular cloud regions by Fukui et al. (2014), who argued for a significant amount of cold HI gas hidden behind optically thick HI as well as dark clouds. 

\subsection{High $\Hfactor$ correction in the GC direction}

High correction factor of $\Hfactor \sim 3.6\pm 0.8$ was obtained close to the GC at $|l|\le 5\deg$, and $3.4\pm 0.9$ at $|l|\le 10\deg$.
This high $\Hfactor$ means that the galactic disk between the Sun and the GC is more abundant in HI gas with density by a factor of $\sim 3-4$ times higher than those so far derived for optically-thin assumption. In fact, it has long been thought that HI is deficient in the inner Galaxy at $R\sim 8$ kpc, where molecular gas is dominant over HI (Nakanishi and Sofue 2003, 2006, 2016), but it might not be the case when the finite optical thickness is taken into account.
 
If such high $\Hfactor$ correction is applied to HI density map of the Galaxy (Nakanishi and Sofue 2003; 2005; 2015), we may need to reconsider the ISM composition in the Galaxy. It may increase the HI mass by several times in the inner Galaxy. It will affect the ISM physics such as the study of molecular fraction, and the HI to H$_2$ phase transition theory as well as to the star formation efficiently via the Schmidt law (Sofue and Nakanishi 2016, 2017b). Since the dynamical mass distribution has been independently determined using the rotation curve, the fractional/abundance consideration of the stellar vs ISM components in the inner Galaxy may be revised by such increase in the HI mass.

\section{Summary and Discussion}

\subsection{Summary} 

We derived a formula to calculate the ratio $\Hfactor$ of volume density of HI gas with finite optical thickness to that under assumption of low opacity. The ratio can be used as a correction factor to estimate the HI densities from currently derived values for optically thin assumption. We presented a map of the $\Hfactor$ value on the sky and a corrected map of the local HI density $n$ in the solar vicinity. 

In order to apply the method, we determined the spin temperature $\Ts$ using the saturated brightness in VDR toward the GC and a local arm, and obtained $\Ts\simeq 146.2\pm 16.1$ K and $144.4\pm 6.8$ K, respectively. We also applied the least $\chi^2$ fitting to the $\Tb$ distribution in the Galactic plane at $|l|\le 30\deg$ around the VDR. We obtained the best-fit set of $\Ts$ and $n$ to be $\Ts=146.8\pm 10.7$ K and $n=0.89\pm 0.14$. Considering these values, we adopted $\Ts=146$ K in the analysis for $\Hfactor$. 

The local values of the correction factor was obtained to be $\Hfactor \sim 1.2$ in the off-Galactic plane region, and $\Hfactor \sim 1.3$ in the inter VDR (off-solar circle) directions in the Galactic plane. Thus, the local HI density, e.g. within 100-200 pc around the Sun, will be larger by a factor of $\sim 1.2$ than the currently estimated density.

A modest value of $\Hfactor \sim 2$ was obtained in the VDR toward local arms (solar-circle) and anti-GC directions. These indicate higher HI mass in the spiral arms and galactic disc at $R\sim 8$ kpc by a factor of $\sim 2$ than the current density estimations. 

A value as high as $\Hfactor\sim 3.6\pm 0.8$ was obtained in the VDR toward GC at $|l|\le 5\deg$. This means that HI density, and therefore the mass, in the inner Galaxy is higher by a factor of $\sim 3-4$ than that currently estimated density and mass under the assumption that the HI line is optically thin.
 
\subsection{Limitation of the method}
The presently developed method to determine the HI spin temperature was applied to the saturated HI brightness region at $\vlsr=0$ \kms in the VDR as shown in figure \ref{VDR}. Although we showed the result for $\vlsr=0$ \kms as the typical example, this method is also possible to be applied to other velocities, if the HI brightness is saturated in the considered region.

Using the spin temperature, the brightness temperature is related to the volume density through equations (\ref{nthick}) and (\ref{nthin}), because the term $dv/dx$ is a known quantity from the rotation curve. Thus, the column density $N$, which is an integration of spectral distribution of $\Tb$ in a finite velocity range, is not used in the present analysis. In this sense, the present method is not a general method to measure  {typical} HI clouds, but is a specific method for the VDR in the Galactic disk with  {the} known kinematics as shown in figure \ref{VDR}.  

\subsection{WNM shadowed by CNM }

We have so far assumed that the HI gas is composed of a single component either of CNM or WNM, and that the spin temperature is constant in each component. However, the real ISM would be a mixture of both components (Field et al. 1969).

If the HI column is low and the CNM is optically thin, both the WNM and CNM emissions are observed simultaneously. This may in fact happen in inter-VDR directions, where the traditional use of equation (\ref{eqNthin}) is valid regardless the spin temperature, giving reasonable estimation of the HI column density, because the equation does not include the spin temperature.

If the column density is so large that the CNM is optically thick, HI line emission from the WNM is significantly absorbed, and the observed brightness temperature becomes saturated around the spin temperature of CNM. In this case, the emission from the WNM is shadowed by the CNM.  

Thus, the WNM on the same line of sight is difficult to be detected, being shadowed by the CNM. However, since the CNM and WNM are in thermal pressure equilibrium (Field et al. 1969), the density of WNM is smaller than that of CNM by a factor of $\Ts ({\rm CNM})/\Ts ({\rm WNM}) \sim 146 {\rm K}/(2000 - 10^4){\rm K}\sim 0.07-0.01$. The mass of WNM on the same sight line would be, therefore, smaller than CNM by  {a factor of 10 to 100}. Hence, the correction for the CNM density and mass is a second order problem, and does not affect the HI mass estimation significantly.

\subsection{Uncertainty of $\Ts$ and propagation into $\Gamma$}

We have assumed a constant spin temperature of $\Ts=146$ K according to the determination in section \ref{secTs}. We here remember that the true spin temperature includes uncertainties arising from:
\begin{itemize}
\item Mixture effect of the CNM and WNM that have quite different spin temperatures, while the effect was shown to be somehow suppressed by the shadowing by CNM, as above;
\item Variation due to intrinsic dispersion in inhomogeneous interstellar medium composed of diffuse and dense cloud HI, as well as colder molecular clouds;
\item Larger scale variation in the Galactic disk as a function of the galacto-centric distance.
\end{itemize}
Although the here derived temperature already includes all these during the determination as an average, there might be some systematic effects adding further uncertainty. We here consider how such uncertainty propagates by giving artificial errors as $\delta \Ts=5, \ 10,$ and 15 K.

Besides $\Ts$, continuum background brightness $\Tc$ estimated from the 2.3 GHz survey may include some systematic error such as due to incorrect spectral index and/or ratio of thermal to nonthermal emissions during the conversion from 2.3 to 1.4 GHz brightness. This effect would be on the order of a few K in the strongest region in the inner Galaxy, and would not be crucial in the present study within the estimation errors. For a more accurate analysis, this problem can be solved by using real 1.4 GHz continuum data at comparable resolutions, while it is beyond the scope of this paper.

We now consider the propagation of $\delta \Ts$ into $\Gamma$, assuming that $\Tc=0$ K. When the gas is optically thin, $\Ts$ does not appear in the conversion relation from HI intensity to density, and hence $\delta \Ts$ does not affect $\Gamma$.
If the gas is optically grey or thick, $\delta \Ts$ included in the non-linear term propagates to the error in the correction factor, $\delta \Gamma$. For $\Tc \ll \Ts$, the propagation is approximated by
\begin{equation}
{\delta \Gamma \over \Gamma} \simeq {\delta \Ts \over \Ts }
\left( {1 \over \Gamma (1-\Tb/\Ts)} -1\right).
\end{equation} . 

Figure \ref{GamError} shows the variation of $\delta \Gamma$ as a function of $\Tb/\Ts$ for $\delta \Ts=5,\ 10$ and 15 K. It is readily seen that $\delta \Hfactor$ is small at $\Tb< \sim 0.8 \Ts$, or at $\Tb < \sim 120$ K for $\Ts=146$ K. Although $\delta \Gamma$ increases steeply beyond, it is still sufficiently small when $\delta \Ts$ is less than $\sim 10$ K.

If $\delta \Ts$ is larger, e.g. $>\sim 15$ K, $\delta \Gamma$ becomes comparable or greater than $\Gamma$ at $\Tb>\sim 140$ K. This will be crucial only in the VDR toward the GC, where we found high $\Gamma $ and $\Tb$. However, considering the procedure of determining $\Ts$ in section \ref{secTs}, such a large uncertainty should not be present. We may, therefore, consider that the present measurements may not be changed largely within the errors.

	\begin{figure} 
\begin{center}  
\includegraphics[width=7cm]{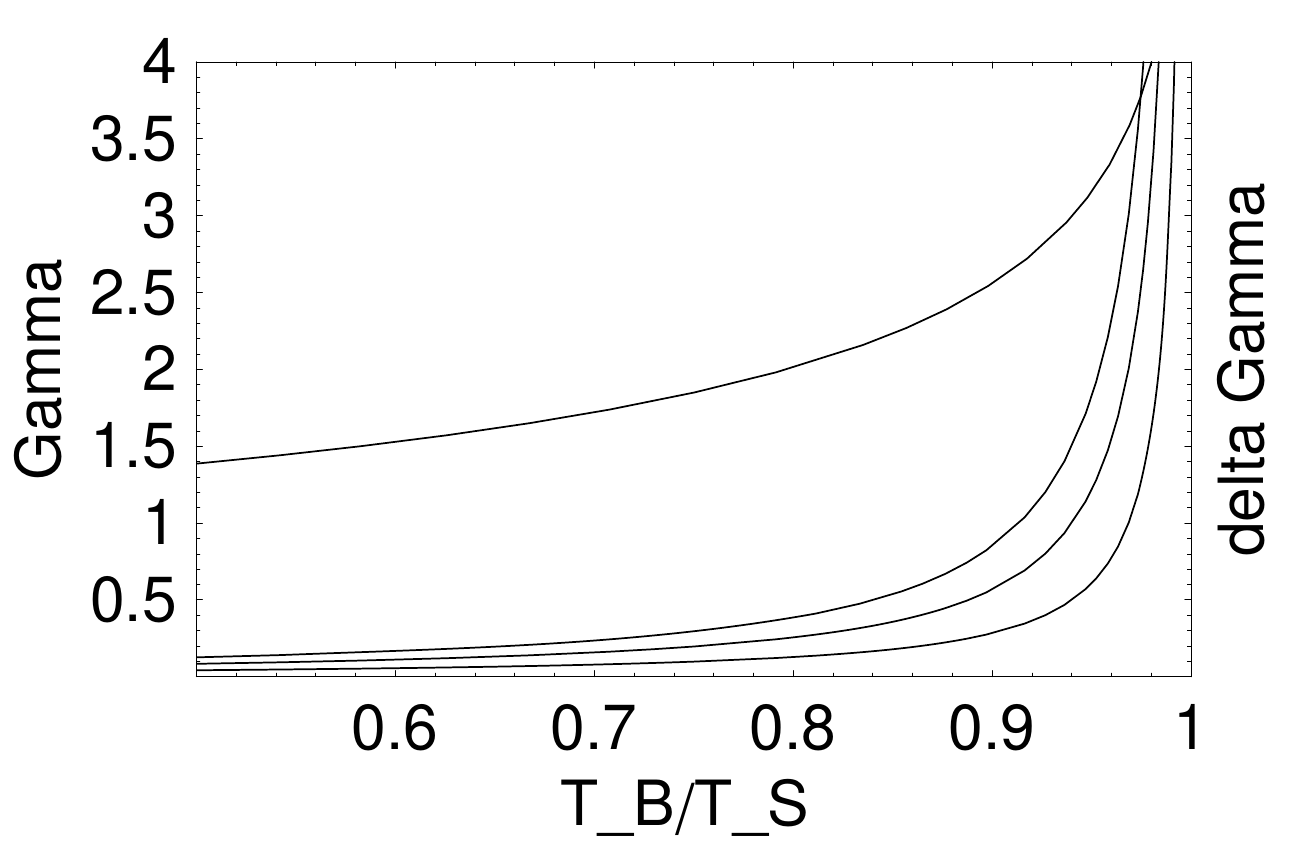}     
\end{center}
\caption{The error, $\delta \Hfactor$, in the correction factor propagating from the uncertainty, $\delta \Ts$, of spin temperature, plotted against $\Tb/\Ts$ for three cases of $\delta \Ts=5,\ 10,$ and 15 K (from bottom to up) and $\Ts=146$ K. The upper-most curve shows $\Gamma$.  }  
\label{GamError} 
	\end{figure}


\begin{thebibliography}{}   

\bibitem[Brown et al.(2014)]{2014ApJS..211...29B} Brown, C., Dickey, J.~M., 
Dawson, J.~R., \& McClure-Griffiths, N.~M.\ 2014, ApJS, 211, 29  

\bibitem[\protect\citeauthoryear{Chengalur, Kanekar, \& Roy}{2013}]{2013MNRAS.432.3074C} Chengalur J.~N., Kanekar N., Roy N., 2013, MNRAS, 432, 3074   
\bibitem[Dickey et al.(2003)]{2003ApJ...585..801D} Dickey, J.~M., McClure-Griffiths, N.~M., Gaensler, B.~M., \& Green, A.~J.\ 2003, ApJ, 585, 801 


\bibitem[\protect\citeauthoryear{Field, Goldsmith, \& Habing}{1969}]{1969ApJ...155L.149F} Field G.~B., Goldsmith D.~W., Habing H.~J., 1969, ApJ, 155, L149 



\bibitem[Fukui et al.(2014)]{2014ApJ...796...59F} Fukui, Y., Okamoto, R., Kaji, R., et al.\ 2014, ApJ, 796, 59 

\bibitem[Fukui et al.(2015)]{2015ApJ...798....6F} Fukui, Y., Torii, K., 
Onishi, T., et al.\ 2015, ApJ, 798, 6 




\bibitem[\protect\citeauthoryear{Li \& Goldsmith}{2003}]{2003ApJ...585..823L} Li D., Goldsmith P.~F., 2003, ApJ, 585, 823  .

\bibitem[\protect\citeauthoryear{Goldsmith \& Li}{2005}]{2005ApJ...622..938G} Goldsmith P.~F., Li D., 2005, ApJ, 622, 938 

\bibitem[\protect\citeauthoryear{Heiles \& Troland}{2003}]{2003ApJ...586.1067H} Heiles C., Troland T.~H., 2003a, ApJ, 586, 1067  

\bibitem[\protect\citeauthoryear{Heiles \& Troland}{2003}]{2003ApJS..145..329H} Heiles C., Troland T.~H., 2003b, ApJS, 145, 329  

\bibitem[\protect\citeauthoryear{Hou \& Han}{2014}]{2014A&A...569A.125H} Hou L.~G., Han J.~L., 2014, A\&A, 569, A125  

\bibitem[Jonas et al.(1985)]{1985A&AS...62..105J} Jonas, J.~L., de Jager, G., \& Baart, E.~E.\ 1985, AA Suppl, 62, 105 
 
\r Li, D., Goldsmith, P. F. 2003 ApJ 585, 823.

\bibitem[Kalberla et al.(1985)]{1985A&A...144...27K} Kalberla, P.~M.~W., Schwarz, U.~J., \& Goss, W.~M.\ 1985, AA 144, 27 

\r Kalberla, P. M. W., Burton, W. B., Hartmann, D., et al. 2005, AA, 440, 775  

\r Kuchar, T. A., Bania, T. M. 1990 ApJ 352, 192. 

\bibitem[\protect\citeauthoryear{Large, Mathewson, \& Haslam}{1961}]{1961MNRAS.123..123L} Large M.~I., Mathewson D.~S., Haslam C.~G.~T., 1961, MNRAS, 123, 123 
  

\bibitem[\protect\citeauthoryear{Liszt}{1983}]{1983ApJ...275..163L} Liszt H.~S., 1983, ApJ, 275, 163  

\bibitem[\protect\citeauthoryear{Liszt}{2001}]{2001A&A...371..698L} Liszt H. S., 2001, A\&A, 371, 698   
 
\bibitem[\protect\citeauthoryear{Liszt, Braun, \& Greisen}{1993}]{1993AJ....106.2349L} Liszt H.~S., Braun R., Greisen E.~W., 1993, AJ, 106, 2349 
 

\bibitem[\protect\citeauthoryear{Mebold et al.}{1982}]{1982A&A...115..223M} Mebold U., Winnberg A., Kalberla P.~M.~W., Goss W.~M., 1982, A\&A, 115, 223 
 

\bibitem[\protect\citeauthoryear{Miyamoto \& Nagai}{1975}]{1975PASJ...27..533M} Miyamoto M., Nagai R., 1975, PASJ, 27, 533  

\bibitem[Murray et al.(2015)]{2015ApJ...804...89M} Murray, C.~E., Stanimirovi{\'c}, S., Goss, W.~M., et al.\ 2015, ApJ, 804, 89 

\r Nakanishi, H., Sofue, Y.\ 2006, PASJ, 58, 847 
\r Nakanishi, H., Sofue, Y.\ 2003, PASJ, 55, 191   
\r Nakanishi, H., Sofue, Y.\ 2016, PASJ, 68, 5  
\r Roberts, D. A., Goss, W. M., Kalberla, P. M. W., Herbstmeier, U., Schwartz, J. J. 1993 AA, 274, 427. 

\bibitem[\protect\citeauthoryear{Roy et al.}{2013}]{2013MNRAS.436.2352R} Roy N., Kanekar N., Braun R., Chengalur J.~N., 2013a, MNRAS, 436, 2352  

\bibitem[\protect\citeauthoryear{Roy, Kanekar, \& Chengalur}{2013}]{2013MNRAS.436.2366R} Roy N., Kanekar N., Chengalur J.~N., 2013b, MNRAS, 436, 2366  


\bibitem[\protect\citeauthoryear{Sofue \& Nakanishi}{2016}]{2016PASJ...68...63S} Sofue Y., Nakanishi H., 2016, PASJ, 68, 63 


\bibitem[\protect\citeauthoryear{Sofue \& Nakanishi}{2017}]{2017MNRAS.464..783S} Sofue Y., Nakanishi H., 2017a, MNRAS, 464, 783 

\bibitem[\protect\citeauthoryear{Sofue \& Nakanishi}{2017}]{2016arXiv161005396S} Sofue Y., Nakanishi H., 2017b PASJ in press, arXiv:1610.05396 
 
\bibitem[Sofue 
\& Reich(1979)]{1979A&AS...38..251S} Sofue, Y., \& Reich, W.\ 1979, AA Suppl, 38, 251 

\bibitem[\protect\citeauthoryear{Stark et al.}{1994}]{1994A&A...281..199S} Stark R., Dickey J.~M., Burton W.~B., Wennmacher A., 1994, A\&A, 281, 199 
 

\bibitem[\protect\citeauthoryear{Wolfire et al.}{1995}]{1995ApJ...443..152W} Wolfire M.~G., Hollenbach D., McKee C.~F., Tielens A.~G.~G.~M., Bakes E.~L.~O., 1995, ApJ, 443, 152 
 



\end{thebibliography}
\end{document}